\documentclass[aps,prb,twocolumn,showpacs]{revtex4-1}

\usepackage{graphicx} 
\usepackage{amssymb}
\usepackage{color}
\usepackage{amsmath}
\usepackage{hyphenat}
\usepackage{hyperref}

\newcommand{\ve}[1]{\mathbf{#1}}

\begin{document}

\title{Orbital tomography of hybridized and dispersing molecular overlayers}

\author{Thomas Ules}
\author{Daniel L\"uftner}
\author{Eva Maria Reinisch}
\author{Georg Koller}
\author{Peter Puschnig}\email{peter.puschnig@uni-graz.at}
\author{Michael G. Ramsey}\email{michael.ramsey@uni-graz.at}
\affiliation{University of Graz, Institute of Physics, NAWI Graz, Universit\"atsplatz 5, 8010 Graz, Austria}

\date{\today}

\begin{abstract}
With angle resolved photoemission experiments and \emph{ab-initio} electronic structure calculations, the pentacene monolayers on Ag(110) and Cu(110) are compared and contrasted allowing the molecular orientation and an unambiguous assignment of emissions to specific orbitals to be made.
On Ag(110), the orbitals remain essentially isolated-molecule like, while strong substrate-enhanced dispersion and orbital modification are observed upon adsorption on Cu(110).
We show how the photoemission intensity of extended systems can be simulated and that it behaves essentially like that of the isolated molecule modulated by the band dispersion due to intermolecular interactions.
\end{abstract}



\maketitle

\section{Introduction}

The effects arising upon adsorption of $\pi$ conjugated molecules on metal substrates are of great interest due to their importance in organic electronics. The ability to identify electronic states with specific molecular orbitals and to determine their energy ordering is vital to the understanding of overlayer/substrate systems. This can be particularly difficult for strong chemisorptive interactions where significant broadening or level splitting occurs or when intermolecular dispersing bands are formed. In the last decade \emph{ab-initio} electronic structure calculations, particulary within density functional theory (DFT), have become almost indispensable in the interpretation of experimental results.
However, it is becoming recognized that DFT results can be misleading owing to approximations for exchange-correlation effects which may severely affect predicted adsorption geometries and/or the electronic structure of organic/metal interfaces \cite{Puschnig2011}.
 The pentacene/Cu interface is such a case. Despite numerous experimental and theoretical studies, there is as yet no consensus on the orbital assignment. Here we demonstrate how the angle resolved photoemission (ARUPS) technique, that is becoming known as orbital tomography \cite{Ziroff2010,Puschnig2011,Stadtmuller2012a,Wiessner2012a,Reinisch2013,Luftner2013,Feyer2014}, can provide a definitive assignment of the emissions even for strongly interacting extended two-dimensional (2D) systems and give insight into the nature of dispersion and hybridization.

The pentacene (5A) monolayer on Cu(110) has been extensively studied. From their ARUPS investigations Seki et al.~\cite{Yamane2007,Yamane2008} concluded that selection rule arguments cannot explain the photoemission behavior and suggested strong hybridization with the substrate could be modifying the orbital symmetry \cite{Yamane2008}. They also suggested the appearance of dispersing, interface-induced states arising from substrate interactions \cite{Yamane2008}. Ferretti et al.~\cite{Baldacchini2007,Ferretti2007} introduced the possibility that the mixing of 5A molecular orbitals with the Cu substrate leads to ''interaction states localized at the interface'', where their calculations suggest partial occupation of the lowest unoccupied molecular orbital (LUMO) whose symmetry might not be directly related to the original molecular state. 
On the same Cu(119) vicinal surface, scanning tunneling microscopy (STM) images also suggest partial occupation of the LUMO and dispersive electronic states associated with a perturbed
electron charge density distribution. \cite{Annese2008} 
 Also a  recent combined DFT and ARUPS study  of the 5A/Cu(110) system concludes that only partial LUMO occupation takes place \cite{Muller2012}. 

The simple relationship between the angular distribution of the photoemission current and the Fourier transform of the emitting molecular orbital has been shown to be reasonable for a number of molecular adsorbate systems on various noble metal surfaces \cite{Puschnig2009a,Ziroff2010,Puschnig2011,Luftner2013,Feyer2014}. This allows molecular orientations to be determined \cite{Puschnig2009a,Feyer2014}, molecular orbital energy ordering to be deduced \cite{Puschnig2011,Willenbockel2012} and even the reconstruction of the molecular orbitals in real space \cite{Luftner2013}. For these systems, the angular-dependent emissions can essentially be accounted for by the photoemission from isolated molecules, thus intermolecular orbital overlap plays a minor role. However, in extended 2D overlayer systems, a description in terms of isolated molecular orbitals is no longer strictly appropriate. Here, with the comparison between pentacene monolayers on Ag(110) and Cu(110), we show how orbital tomography and the Fourier transform description can be applied to extended systems with strongly dispersing emissions. In so doing, we provide a definitive description of the 5A/Cu ARUPS and show that the LUMO is in fact fully occupied and displays a substrate-induced dispersion which is significantly larger than that reported for similar organic overlayer systems \cite{Wiessner2012b,Wiessner2013}. Moreover, unlike all previous tomography studies we here show evidence for a modification of orbital shape on adsorption.

\section{Experimental Details}

\begin{figure}[!htb]
	\includegraphics[width=0.7\columnwidth]{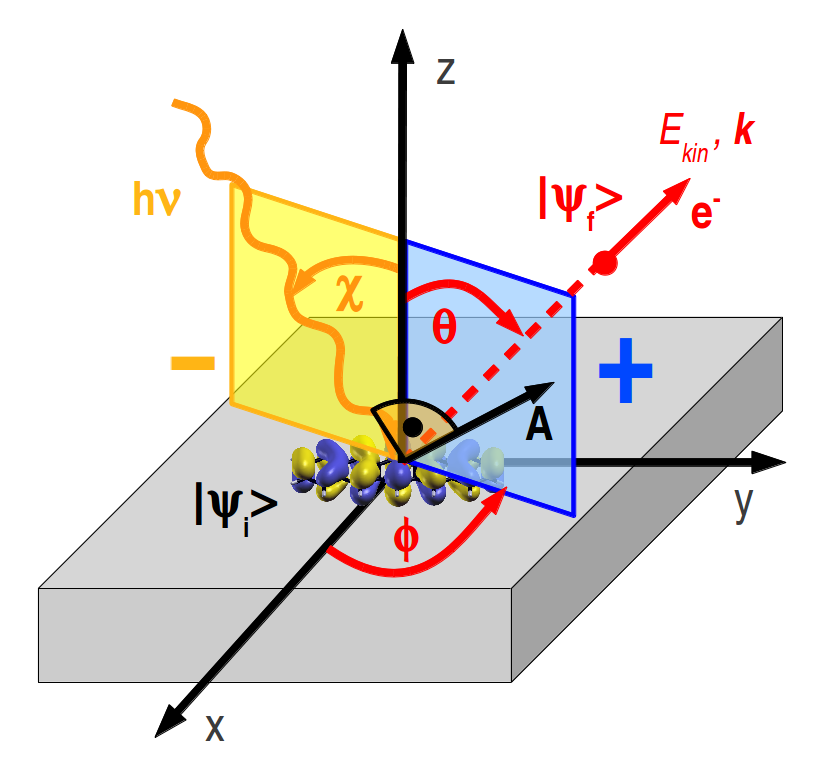}
	\caption{\label{FigS1} Sketch of an ARUPS experiment declaring the geometry of the experimental setup. The incoming photon with the energy $h\nu$, the incidence angle $ \chi $ and vector potential \textbf{A} excites an electron from the initial state $ \psi_i $ to the final state $ \psi_f $. This final state is characterized by the kinetic energy $ E_{\mathrm{kin}} $ and the momentum vector \textbf{k} and the outgoing photoelectron is detected as a function of $ E_{\mathrm{kin}} $ and emission direction, defined by the polar angle $ \theta $ and the azimuthal angle $ \phi $. Where forward emissions are denoted for by a positive parallel momentum and backward by a negative, respectively.}
\end{figure}

Photoemission experiments were performed at BESSYII using a toroidal electron-energy analyzer described previously \cite{Broekman2005} which was attached to the beamline U125/2-SGM of the synchrotron radiation facility BESSY II, Helmholtz-Zentrum-Berlin. Photon energies of 30 and 35 eV and an incidence angle of $ \chi $ = 40$^\circ$ with respect to the surface normal were used. The polarization direction is in the specular plane, which also contains the photoelectron trajectory measured. Emitted photoelectrons are recorded simultaneously with polar angles $\theta$ of $-80^\circ$ to $+80^\circ$ with respect to the surface normal in an energy window of 1 eV. The energy window is divided into 40 individual slices which goes well below the analyzer’s intrinsic energy resolution of 150 meV. Note, however, for the presented momentum maps we take only the data from the side of the emission direction that is pointing in the orientation of the electric field vector of the incident photons, this is referred to as the positive side (see Fig.~\ref{FigS1}), i.e., $\theta$ = 0$^\circ$ to +80$^\circ$. This maximizes the polarization factor appearing in the photocurrent cross section (see Eq.~\ref{eq4}). It is noted that the molecular features are enhanced relative to the substrate emissions on the positive side (compare Fig.~\ref{Fig2}c). To obtain the full ( k$_{x}$, k$_{y}$ ) range for the presented momentum maps at constant binding energy, azimuthal scans are made by rotating the sample around the surface normal in 1$^\circ$ steps for an azimuthal angle range $>180^\circ$ and then imposing the substrate’s twofold symmetry to obtain the full 360$^\circ$. The angular emission data are then converted to parallel momentum components k$_{x}$ and k$_{y}$ using the relation
\begin{eqnarray}
k_x = \sqrt{\frac{2m_e}{\hbar^2}E_{\mathrm{kin}}}\sin\theta \cos\phi
\label{eq1}
\end{eqnarray}

\begin{eqnarray}
k_y = \sqrt{\frac{2m_e}{\hbar^2}E_{\mathrm{kin}}}\sin\theta \sin\phi ,
\label{eq2}
\end{eqnarray}
to create the momentum maps.
The Cu(110) and Ag(110) substrates were prepared in the conventional way by a sequence of sputter-annealing cycles. 5A molecules were evaporated from an effusion cell onto the surfaces at room temperature with the amount monitored by a microbalance. The resulting monolayer LEED structures are $\left( \begin{array}{cc} 3 & -1 \\ 1 & 4 \end{array} \right)$ for pentacene on Ag(110) and $\left( \begin{array}{cc} 6.5 & -1 \\ -0.5 & 2 \end{array} \right)$ for pentacene on Cu(110). In both cases mild annealing improved the order. On Cu(110) the monolayer was annealed at 200$^\circ$C which is above the 5A sublimation temperature whereas on Ag(110) the molecule substrate bond is weaker and the annealing temperature must not exceed 140$^\circ$C so as not to desorb from the monolayer.

\section{Computational Details}

\subsection{Density functional calculations}

All theoretical results presented here are obtained within the framework of density functional theory (DFT) using the VASP code \cite{Kresse1993,Kresse1999}.  We have performed three types of calculations: firstly for the isolated pentacene molecule, secondly for a two-dimensional, extended  free-standing layer of pentacene, and thirdly for monolayers of pentacene adsorbed on Ag(110) and Cu(110) surfaces.

The isolated molecule calculations were performed using a supercell with a minimum of 15~{\AA} vacuum between pentacene's periodic replica. We use the generalized gradient approximation (GGA) \cite{Perdew1996} for exchange-correlation effects, and the projector augmented wave (PAW) method \cite{Bloechl1994}. The simulated momentum maps of the pentacene HOMO and LUMO orbitals shown in Fig.~\ref{Fig1} are obtained as Fourier transforms of the respective Kohn-Sham orbitals as described previously \cite{Puschnig2009a}.

Electronic structure calculations for the freestanding monolayer of 5A have been carried out using the repeated slab approach with a vacuum layer of  20~{\AA} between adjacent layers. The generalized gradient approximation (GGA) \cite{Perdew1996} is used for exchange-correlation effects, and the projector augmented waves (PAW) \cite{Bloechl1994} approach was used allowing for a relatively low kinetic energy cut-off of about 400 eV. We use a Monkhorst-Pack  $6 \times 12 \times 1$ grid of $k$-points \cite{Monkhorst1976}, and a first-order Methfessel-Paxton  smearing of 0.1 eV \cite{Methfessel1989}. 
The simulation of momentum maps for extended systems will be described in the  subsequent section.

Finally, we have also performed DFT calculations of pentacene monolayers adsorbed on Ag(110) and Cu(110). The substrate is taken into account within the repeated slab approach by using five metallic layers with an additional vacuum layer of  15~{\AA} between slabs. To avoid spurious electrical fields, a dipole layer is inserted in the vacuum region \cite{Neugebauer1992}. 
In the case of 5A/Ag(110), we have taken into account the experimental LEED structure $\left( \begin{array}{cc} 3 & -1 \\ 1 & 4 \end{array} \right)$ mentioned above and relaxed the atomic positions of the molecule and the first metallic layer considering van-der-Waals interactions by employing the empirical correction scheme according to Grimme \cite{Grimme2006}. Exchange correlation effects were treated either within the GGA \cite{Perdew1996} or within the  hybrid functional HSE \cite{Heyd2006} with k-point meshes of $9 \times 6 \times 1$ and $6 \times 4 \times 1$, respectively, and a first-order Methfessel-Paxton  smearing of 0.1 eV.
For 5A/Cu(110), we have chosen a commensurate structure  $\left( \begin{array}{cc} 6 & 0 \\ 0 & 2 \end{array} \right)$ and an adsorption site similar to a previous study \cite{Muller2012}. Also, here the electronic structure is calculated within the GGA and HSE using k-point meshes of $9 \times 12 \times 1$ and $4 \times 6 \times 1$, respectively, and using a first-order Methfessel-Paxton  smearing of 0.1 eV.



\subsection{Simulation of ARUPS maps}

The Kohn-Sham energies $\varepsilon_{n\ve{q}}$ and orbitals $\psi_{n\ve{q}}$ of the relaxed structures serve as input for the subsequent simulation of ARPES intensity maps within the one-step model of photoemission \cite{Feibelman1974}. Here, the angle-resolved photoemission intensity $I(\theta,\phi;E_{\mathrm{kin}},\omega)$ is a function of the azimuthal and polar angles $\theta$ and $\phi$, respectively, the kinetic energy of the emitted electron $E_{\mathrm{kin}}$, and the energy $\omega$ and polarization $\ve{A}$ of the incoming photon:
\begin{eqnarray}
I(\theta,\phi,E_{\mathrm{kin}};\omega,\ve{A}) & \approx & 
\sum_{n}^\textrm{occ} \sum_{\ve{q}}^\textrm{BZ}
\left|\langle \psi_f(\theta,\phi;E_{\mathrm{kin}})|\ve{A}\cdot\ve{p}|\psi_{n\ve{q}}\rangle \right|^2 \nonumber \\ 
& \times & \delta(\varepsilon_{n\ve{q}} + \Phi + E_{\mathrm{kin}} - \omega),
\label{eq3}
\end{eqnarray}
This formula can be viewed as a Fermi's golden rule expression, in which the photocurrent $I$ results from a summation over all occupied initial states $\psi_{n\ve{q}}$, characterized by the band index $n$ and Bloch vector $\ve{q}$, weighted by the transition probability between the initial state and a final state. For the transition operator $\ve{A}\cdot\ve{p}$, the dipole approximation is assumed, where $\bf{p}$ and $\bf{A}$ are the momentum operator and the vector potential connected to the incoming photon. The $\delta$ function  ensures energy conservation where $\Phi$ denotes the work function.

We further approximate the final state $\psi_f$ by a plane wave \cite{Gadzuk1974}. As outlined in more detail in a previous paper \cite{Puschnig2009a}, and also noted earlier \cite{Shirley1995,Mugarza2003}, this approximation allows us to greatly simplify the evaluation of the matrix element appearing in Eq.~\ref{eq3}. If we denote the wave vector of the of the final, free-electron state by $\ve{k}$, thus $E_{\mathrm{kin}}=\frac{\hbar^2}{2m}k^2$, Eq.~\ref{eq3} simplifies to
\begin{eqnarray}
I(k_x,k_y,E_{\mathrm{kin}};\omega,\ve{A}) & \approx & 
\sum_{n}^\textrm{occ} \sum_{\ve{q}}^\textrm{BZ}
\left| \ve{A} \cdot\ve{k} \right|^2
\left| \langle e^{i \ve{k r}}| \psi_{n\ve{q}}\rangle \right|^2 \nonumber \\ 
& \times & \delta(\varepsilon_{n\ve{q}} + \Phi + E_{\mathrm{kin}} - \omega).
\label{eq4}
\end{eqnarray}
We obtain the  simple result that the matrix element from a given initial state $n\ve{q}$ is proportional to the square modulus of the Fourier transform of the initial state wave function $\psi_{n\ve{q}}$ 
modulated by the weakly angle-dependent factor $|\bf{A}\cdot\bf{k}|^2$ which depends on the angle between the polarization vector $\bf{A}$ of the incoming photon and the direction $\ve{k}$ of the emitted electron.

Note Eq.~\ref{eq4} can be applied to single molecules as well as to extended states such as organic layer adsorbed on metallic surfaces. In the former case the summation of the Brillouin zone (BZ) reduces to just one point, the $\Gamma$ point, while for the latter situation, the possible dispersing bands are taken into account by the band structure $\varepsilon_{n\ve{q}}$ and Bloch states $\psi_{n\ve{q}}$ and an appropriate sampling of the  Brillouin zone.

\section{Results}

\subsection{Momentum maps}

\begin{figure}[!htb]
	\includegraphics[width=0.85\columnwidth]{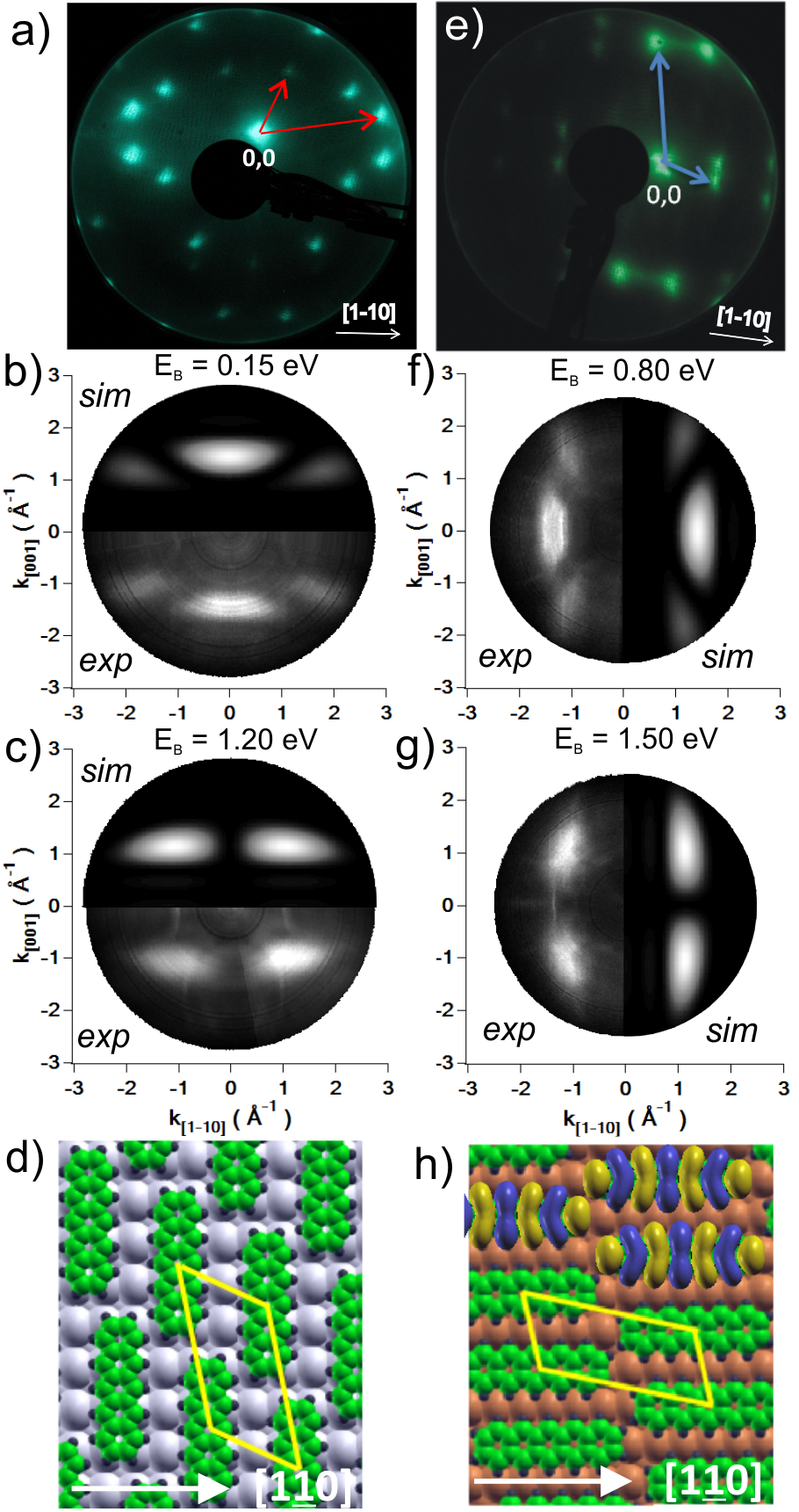}
	\caption{\label{Fig1} Panels a) and e) show LEED images of 5A monolayers on Ag(110) and Cu(110), respectively. b) and c) show ARUPS momentum maps of 5A/Ag(110) at binding energies 0.15 and 1.20 eV (\emph{exp}), respectively, compared to simulated LUMO and HOMO maps of the isolated 5A molecule (\emph{sim}). Panels f) and g) show the corresponding data for 5A/Cu(110) at binding energies of 0.80 and 1.50 eV. Panels d) and h) show structural models of 5A/Ag(110) and 5A/Cu(110), respectively, as deduced from LEED and the momentum maps. In h) the orbital structure of the LUMO is overlaid.}
\end{figure}  

Monolayers of 5A on Ag(110) and Cu(110) were characterized by low energy electron diffraction (LEED) and ARUPS.
Fig.~\ref{Fig1} shows LEED images of the 5A monolayers on the two substrates Ag(110) and Cu(110) as well as the momentum maps taken at the binding energies of the prominent adsorption induced emissions at 0.15 and 1.2 eV on Ag and 0.8 and 1.5 eV on Cu, respectively. 
These maps clearly display the character of the LUMO and HOMO of pentacene as can be seen from the comparison between the measured (\emph{exp}) and the corresponding simulated maps computed for the isolated molecule (\emph{sim}). The simulated LUMO map, for instance, is characterized by a major lobe at (0,1.4) \AA$^{-1}$  and minor lobes at ($\pm1.8$, $1.2$) \AA$^{-1}$.
Indeed the data on Ag has recently been used to reconstruct these orbitals in real space in excellent agreement with calculated orbitals for the isolated molecule \cite{Luftner2013}. 
The maps also unambiguously reveal the \emph{orientation} of the molecules: on Ag(110) the data implies flat lying 5A oriented parallel to the [001] azimuth, while on Cu the molecules orient along the [1-10] azimuth as depicted in  Fig.~\ref{Fig1}d and h, respectively.
It should be noted that the structure and density of the two monolayers are very similar, however, on Ag the molecules lie across while on Cu they lie parallel to the close packed rows of the substrate. 

Thus, the ARUPS momentum maps of Fig.~\ref{Fig1} immediately clarify the 5A/Cu system. In contrast to speculations in the literature, the symmetry of the orbitals is not modified significantly on hybridization. This allows the emissions to be unambiguously assigned. The features at 0.8 and 1.5 eV are the fully occupied LUMO and HOMO and not the HOMO and HOMO-1 emissions as previously and, in the light of our results, erroneously assigned on the basis of a comparison between DFT-calculated and experimental binding energies \cite{Muller2012}. 
On closer inspection of the LUMO map on Cu, however, an internal structure and a change in the $k$-position of the minor lobes becomes visible that will be argued to be the result of intermolecular dispersion and changes in the orbital size, respectively.

\subsection{DFT results for adsorbed monolayers}

\begin{figure}[!htb]
	\includegraphics[width=\columnwidth]{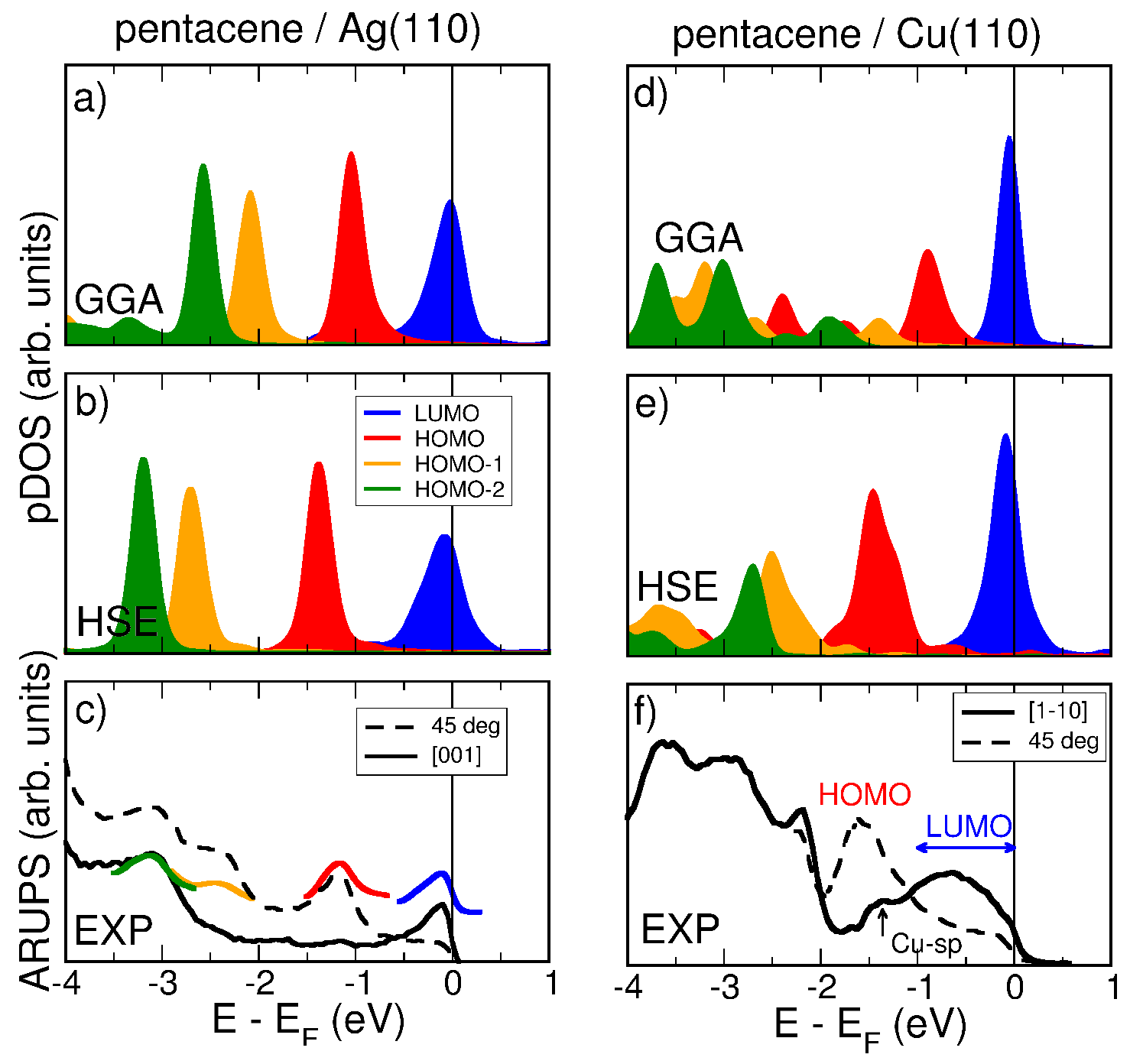}
	\caption{\label{FigS4} Density of states of a monolayer of pentacene on Ag(110) projected onto orbitals of the free pentacene molecule using PBE-GGA (a) and the hybrid functional HSE06 (b). Projections onto the LUMO (blue), the HOMO (red), the HOMO-1 (orange) and the HOMO-2 (green) are shown. For comparison, panel (c) shows experimental ARPES data along the [001] azimuth (black solid line) and an azimuthal direction 45$^\circ$ between [001] and [1-10]. The colored curves are obtained by $k_xk_y$-integrating experimental momentum maps identifying the orbital character of the emissions. Panels (d)--(f) show the corresponding data for a monolayer of pentacene on Cu(110).}
\end{figure}  

Before analyzing this internal structure of the LUMO in more detail, we discuss DFT results of the pentacene monolayers on the Ag(110) and Cu(110) surfaces in comparison with ARPES data. 
Starting from the relaxed structure of 5A/Ag(110), we have computed the density of states projected onto the orbitals of a freestanding pentacene layer, both, within GGA and self-consistently within the hybrid functional HSE \cite{Heyd2004,Heyd2006}. It is known that the incorporation of a fraction of exact exchange in the hybrid HSE mediates self-interaction errors thereby improving the orbital energies and thus the description of adsorbate systems \cite{Korzdorfer2009,Luftner2013a,Luftner2014}.
 The results are shown in panels (a) and (b) of Fig.~\ref{FigS4}. For comparison, panel (c) shows experimental ARPES data recorded over a large energy window along the two different azimuths [001] (black solid line) and in a direction 45$^\circ$ between [001] and [1-10] (black dashed line). Panel (c) also includes the $k_xk_y$-integrated ARPES data of the momentum maps that have been recorded over four energy windows of 1 eV centered around the molecular emissions of the LUMO, HOMO, HOMO-1, and HOMO-2, respectively.
When comparing GGA with HSE results, we observe that the HOMO, HOMO-1 and HOMO-2 features are shifted to somewhat larger binding energies when using the hybrid functional while the position of the LUMO orbital slightly below the Fermi level remains unchanged. Compared to experiment, we note that the HSE calculation clearly improves energy position of the HOMO-1 and HOMO-2 and also leads to a slightly better agreement of the HOMO position. Regarding the LUMO, both functionals GGA and HSE indicate partial occupation in good agreement with experiment.

For the case of Cu(110) (Fig.~\ref{FigS4}, panel d--f), 
HSE again yields much better agreement with experiment. However, the LUMO, of prime importance in this study, is still in poor agreement with experiment. Its computed position, both, obtained within GGA and within HSE, is in error compared to experiment as the LUMO is located at the Fermi level and not fully occupied with only a slight improvement gained by the hybrid functional calculation  \cite{Hofmann2013}. Note that various commensurate structures and adsorption sites did not change this finding. On the other hand, HSE does considerably improve the energy position of the deeper lying orbitals, for instance, the HOMO. As a side-note: It was exactly the wrong energy position of the HOMO in the GGA result which had motivated M\"uller et al. to erroneously assign the LUMO emission to the HOMO. Note that the computed DOS of their paper \cite{Muller2012} agrees with our GGA-DOS and their  experimental ARUPS data is in line with our experiment.

\subsection{1D Band dispersion}

\begin{figure}[!htb]
	\includegraphics[width=\columnwidth]{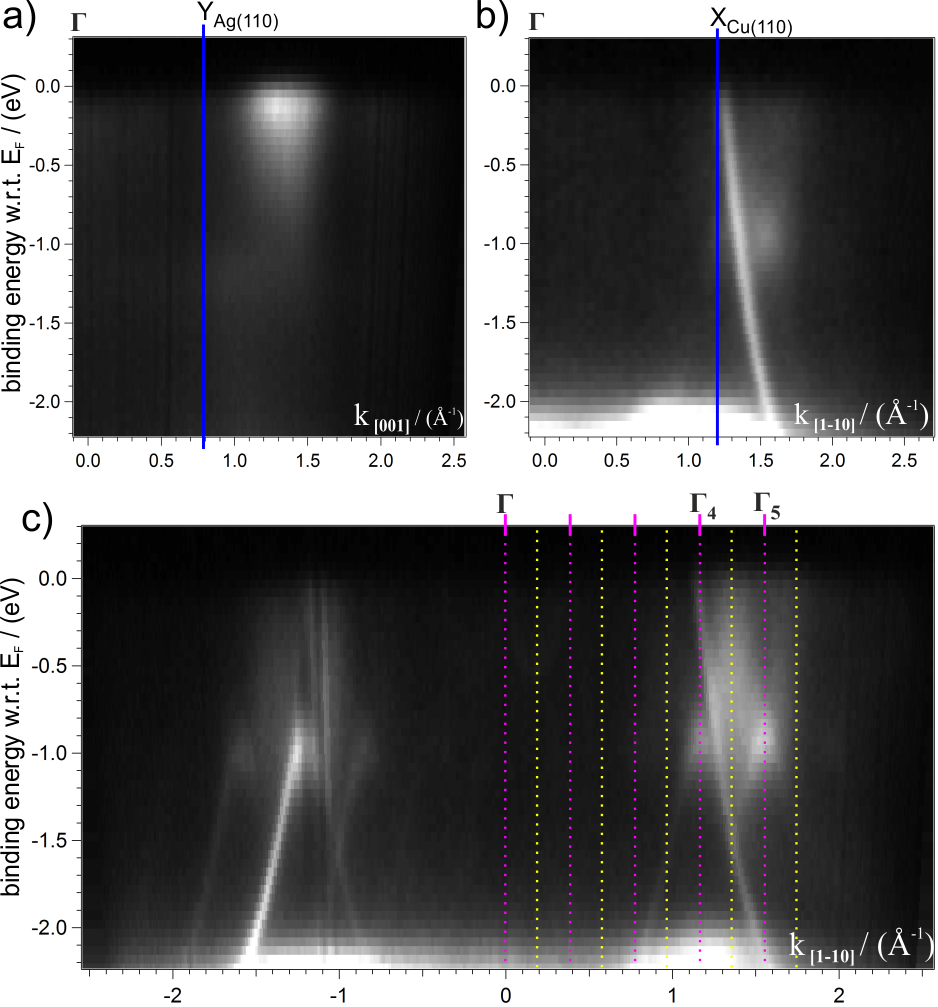}
	\caption{\label{Fig2}a) Bandmap of a 5A monolayer on Ag(110) in the [001] azimuth.  b) and c) show band maps in the [1-10] azimuth of a \emph{half} and a \emph{complete} monolayer of 5A on Cu(110), respectively. Vertical lines indicate the Brillouin zone boundaries (yellow) and the $\Gamma$ points (purple) of the 5A overlayer.}
\end{figure}

Let us now concentrate on the energy dispersion of the LUMO state as suggested by the internal structure of the momentum maps.
Fig.~\ref{Fig2} shows the $E$ \emph{vs.} $k$ relation (bandmap) along the long molecular axes of 5A for a) a monolayer an Ag(110), b) a half monolayer on Cu(110) and c) a complete monolayer on Cu(110) from the Fermi edge ($E_F$) down to 2.2 eV binding energy.
On Ag, the bandmap shows a feature located at the Fermi edge at a $k$ value around 1.35 \AA$^{-1}$ whose intensity gradually tales off with increasing binding energy indicating that the LUMO is half-filled. Note that in this energy range, there are no apparent changes in the momentum maps.
In Fig.~\ref{Fig2}b, the half monolayer of 5A on Cu(110) is investigated. By examining  such a low coverage, with no long-range order apparent from LEED, any effects of intermolecular dispersion are minimized and only effects due to the molecule-substrate interaction are expected. 
The bandmap of the 5A submonolayer on Cu(110) in the [1-10] direction (Fig.~2b), \emph{i.e.}~along the long molecular axes, shows, besides the Cu $d$  and $sp$ bands, emissions from the major 5A LUMO lobe visible around 1 eV binding energy. 
Comparing with Ag, the LUMO shifts from $E_F$ and being half-occupied to a binding energy of roughly 1 eV and being fully occupied on Cu. This is indicative of the strong bonding interaction of 5A with Cu, nevertheless, the band map shows no obvious sign of intermolecular dispersion.

Upon the formation of the well ordered monolayer on Cu, the situation is changed with new structure appearing in both the band and momentum maps. As evident in Fig.~\ref{Fig2}c, the principle molecular induced emission now oscillates in energy from $E_F$ down to 1.2 eV in the range of $k_{[1-10]}$ from 1.0 to 1.9 \AA$^{-1}$ suggestive of strong intermolecular dispersion (see Fig.~\ref{Fig3}b for an enlarged view).
To emphasize the $E$~vs.~$k$ relation of the LUMO along the long molecular axes direction, the Brillouin zone (BZ) boundaries and $\Gamma$ points of the 5A monolayer structure deduced from LEED are indicated in Fig.~\ref{Fig2}c. Note that without differentiation or other enhancements of the raw data, the large LUMO dispersion with the periodicity of the overlayer is clearly evident in the 4th and 5th zones. The high binding energy side of the band is seen to be at $\Gamma$ and the low binding energy at the zone boundaries. Given the LUMO orbital's topology (Fig.~\ref{Fig1}h), the band dispersion running up from $\Gamma$ to the BZ boundary is qualitatively consistent with a chain of pentacene molecules.

\subsection{2D Band dispersion}

\begin{figure}[!htb]
	\includegraphics[width=\columnwidth]{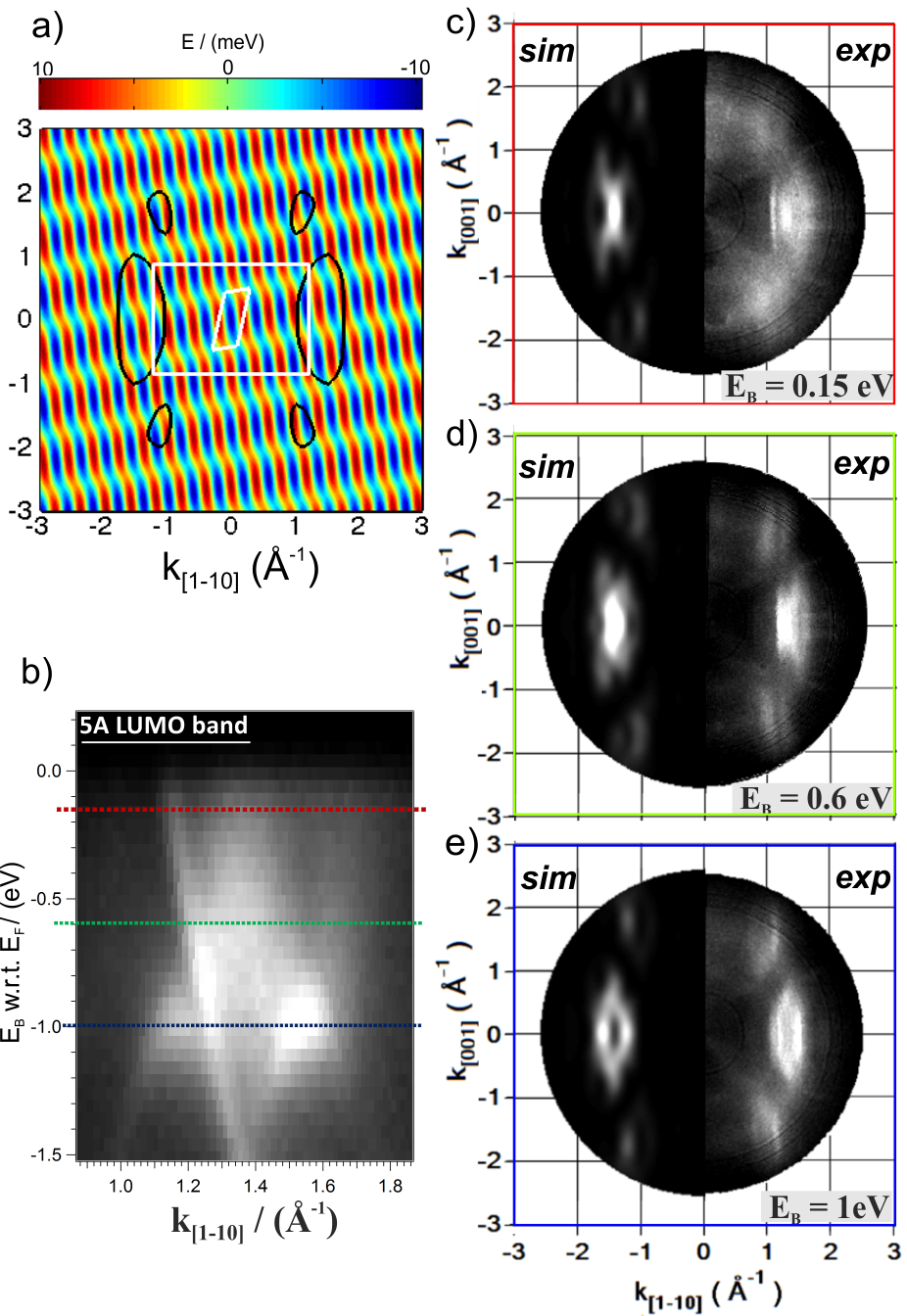}
	\caption{\label{Fig3} a) Color-coded two-dimensional dispersion $E(k_{[001]},k_{[1-10]})$ of the LUMO for a freestanding 5A monolayer with the structure of Fig.~\ref{Fig1}h. Superimposed in white are the Brillouin zones of Cu(110) (rectangle) and the 5A monolayer (rhombus), and in black an intensity iso-line of the calculated isolated 5A LUMO momentum distribution. b) Close-up of the dispersing LUMO band with red, green and blue lines indicating the energy positions at which momentum maps are extracted for comparison with the calculated photoemission momentum distributions in c) top d) middle and e) bottom of the band. Note that in panels c)--e) simulated  momentum maps (\emph{sim}) are compared to experimental maps (\emph{exp}) as detailed in the text.}
\end{figure}

Having qualitatively understood the one-dimensional dispersion along the long molecular axes, a more quantitative and, above all, an understanding of the two-dimensional (2D) LUMO dispersion on Cu and its manifestation in the photoemission experiment is desired. To this end, we first discuss the computed 2D LUMO dispersion of the \emph{free-standing layer} which is plotted in Fig.~\ref{Fig3}a with the band energy $E(k_{[001]},k_{[1-10]})$ color-coded: green indicating the center of the band, and red (blue) the top (bottom) of the band, respectively, where for reasons of clarity just one of the mirror domains is included. Note that although our 5A/Cu(110) HSE calculation including the substrate considerably improves the agreement with experiment over GGA results, we refrained from using it here for the analysis of the experimentally observed band dispersion of the LUMO for the following reason.  The experimental overlayer structure is non-commensurate while, in order to enable a DFT calculation including Cu(110), a commensurate surface unit cell had to be imposed (similar to a previous study \cite{Muller2012}). This difference would result in distinctly different periodicities in the measured and simulated momentum maps preventing a one-to-one comparison. In contrast, the free-standing layer simulation with the correct structure enables a one-to-one analysis of the experimentally observed LUMO dispersion.

Inspection of Fig.~\ref{Fig3}a shows that the character of the dispersion, i.e., the $k$-positions of the tops and bottoms of the band, may well be understood by knowing (i) the respective Brillouin zone, and (ii) the nodal structure of the respective molecular orbital in conjunction with the intermolecular arrangement of molecules. The first determines the reciprocal periodicities and the second the direction of the dispersion, i.e., whether the band would run up or down from $\Gamma$ to the BZ boundary. The main dispersion is along the long real space unit cell vector, \emph{i.e.}, roughly along the long molecular axes. The calculated energy spread of 20 meV for the free-standing layer is, however, very small and more than an order of magnitude lower than the experimentally observed one.

In the ARUPS experiment, one does not observe the 2D dispersion of the LUMO-derived band over the entire $k$-range due to matrix element effects. For isolated molecules, the selection rules are well-described by the FT of the isolated molecular orbitals. 
In analogy, in order to simulate  the photoemission intensity distribution in a quantitative manner, we thus need to consider the wave functions of the extended 2D molecular system and compute their Fourier transforms according to the plane-wave final state approximation \cite{Puschnig2009a}.
Evaluating Eq.~\ref{eq4} for a free-standing layer, we simulate momentum maps (\emph{sim}) at the top, middle and bottom of LUMO-derived band and compare them with the experimental maps (\emph{exp}) taken at the 0.15 eV, 0.60 eV and 1.0 eV binding energy in panels c, d and e of Fig.~\ref{Fig3}, respectively. These energy positions are also indicated as horizontal lines in the enlarged band map shown in panel b. Note that in the simulations, we have taken into account that the monolayer structure consists of two mirror domains. 

The experimental and simulated momentum maps are in remarkably good agreement.
For the major lobe, in going from top to bottom of the band, not only is the general shape in agreement, one also observes an increase in the extension of the feature in $k_{[001]}$ and a splitting of the emission in $k_{[1-10]}$. Although weak there is also agreement in the behavior of the minor lobes. For instance, the shift in $k_{[1-10]}$ from 1.0 to 1.3 \AA$^{-1}$ is seen in both experiment and simulation. Also, in the middle of the band the splitting of the minor lobe seen in the experiment, although difficult to discern, is also observable in the calculated map. We conclude that the photoemission intensity of the extended system is essentially that of the isolated molecule modulated by the intermolecular dispersion. This naturally implies that the practice of searching for dispersion in extended systems outside the $k$-range expected for the isolated molecules is questionable. This is illustrated in the calculated 2D dispersion of Fig.~\ref{Fig3}a by overlaying the isoline of the computed ARUPS map of the the isolated 5A LUMO.

While giving extremely good agreement in the three momentum maps shown in Fig.~\ref{Fig3}c--e, the exact  $E$ vs. $k$ relation is naturally problematic when analyzed in terms of a free-standing layer. 
Indeed, as experience with other systems has shown, the substrate may or may not (depending of the relative energy position of molecular and substrate levels) enhance the dispersion as e.g. for PTCDA/Ag(110) \cite{Wiessner2012b} and NTCDA/Ag(110) or NTCDA/Cu(100) \cite{Wiessner2013}. But in all these cases the \emph{character} of the dispersion is left unaltered and therefore the appearance of momentum maps is also unchanged when going from the free-standing layer to the adsorbed monolayer.  
The \emph{magnitude} of the dispersion is generally influenced by the molecule-substrate interaction.
The case of pentacene/Cu(110) stands out because the observed dispersion is the largest one measured so far and may thus be even termed substrate-\emph{induced} since the band width of the free-standing layer's LUMO band is almost negligible.

Indeed, our calculation for 5A/Cu(110) using the commensurate structure does show enhanced dispersion compared to the free-standing layer without the substrate, however, not as large as the experimental observation presumably as a result of the underestimation of the  LUMO binding energy.



\subsection{Molecule-Metal Hybridization}

\begin{figure}[!htb]
	\includegraphics[width=\columnwidth]{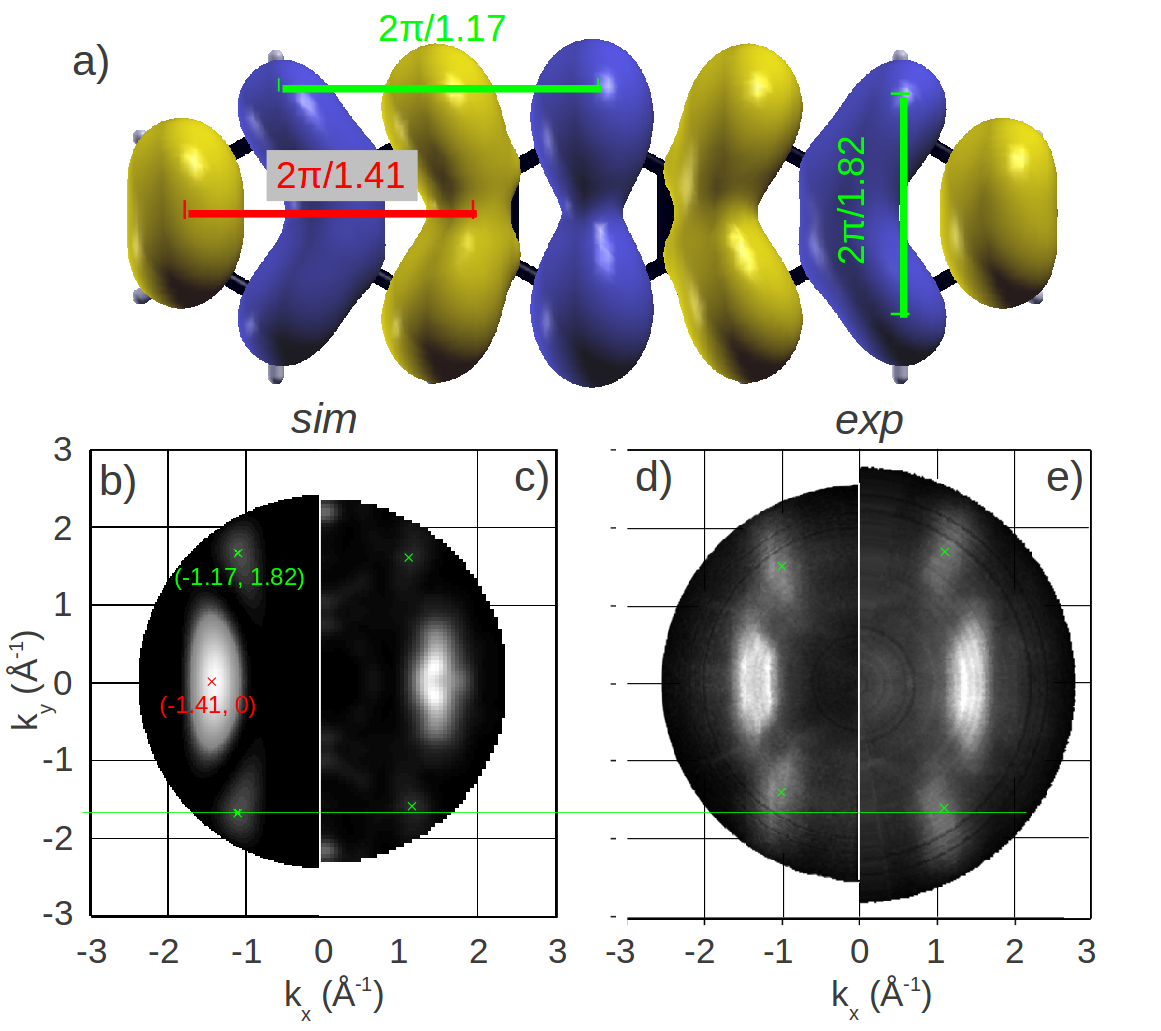}
	\caption{\label{FigS6}a) Calculated LUMO of an isolated 5A molecule in real space, scale bars mark characteristic dimensions as detailed in the text. b) and c) are simulated momentum maps for a free 5A molecule and the monolayer 5A/Cu(110), while d) and e) show experimental ARPES maps of 5A/Cu(110) and 5A/Ag(110), respectively. The red and green crosses indicate the position of the maxima of the LUMO's major and minor lobe, the horizontal green line is a guide for the eye.}
\end{figure}

For the 5A/Cu system, in addition to the structure introduced by dispersion, there is also a distinct change in the $k$-position of the minor lobes in the  $k_{[001]}$ direction. 
In Fig.~\ref{FigS6}, simulated momentum maps of the LUMO for the  gas-phase and for the 5A/Cu(110) adsorbate system are compared to experimental maps of the 5A LUMO on both surfaces.
On inspection of Fig.~\ref{FigS6}d, one sees that on Cu, the minor lobes (green crosses) have  maxima at $k_y = \pm 1.5$~\AA$^{-1}$  which is significantly lower than, both, the simulated values for the isolated molecule (b) and the experimental value on the Ag surface (e) ($\pm 1.8$~\AA$^{-1}$). 
This shift is a consequence of the molecule-copper interaction and can be interpreted as a $\approx 20$\% increase of the lateral orbital size upon adsorption. It can be rationalized by recalling the reciprocal relationships between the LUMO shape in real space and the corresponding momentum space patterns as illustrated in Figs.~\ref{FigS6}a and b. Here the maxima of the major and minor lobes are marked as red and green crosses in the momentum map, respectively, while the corresponding real space dimensions are indicated by scale bars of length $2 \pi/k$. We note that the width of the LUMO orbital along the short molecular axis is reflected by the $k_y$ maximum of the minor lobe.
Thus, unlike all molecular adsorbate systems so far reported \cite{Puschnig2009a,Dauth2011,Luftner2013,Wiessner2012a}, the orbital tomography of the 5A LUMO on Cu(110) shows a distinct modification of the orbital's shape from the isolated molecule. 

Our interpretation of the experimental data in terms of a spatial distortion of the orbital is further supported by our DFT calculations. Starting from the orbital energies and wave functions of the HSE calculation  for 5A/Cu(110), we simulate an ARPES intensity map of the LUMO as shown in Fig.~\ref{FigS6}c. These simulations indeed indicate a shift of the minor lobe peak (green cross) to lower $k$-values upon adsorption on Cu(110). We note, however, that this change is smaller   than the experimental observation. The discrepancy is likely due to a shortcoming of the DFT results regarding the LUMO binding energy (too low) and width (too narrow) at the HSE-DFT level indicating an underestimation of the interaction strength of the LUMO with the substrate. 
 The pronounced difference between the minor lobe peak position of 5A/Cu(110) compared to 5A/Ag(110)   can thus be seen as direct evidence for a strong molecule-metal hybridization for the former.


 While hybridization is generally observed by indirect means in effects on molecular emissions such as energy broadening of orbital emissions \cite{Wiessner2012b} or reduced emission intensity in gaps of bulk states \cite{Berkebile2011}, we here provide direct evidence for changes in orbital shape. Moreover, further signatures of hybridization can be observed as additional Cu-$sp$-like features appearing with the overlayer periodicity (Fig.~\ref{Fig2}). These are best seen on the \emph{negative} $k$ side where the molecular features are weak as shown in Fig.~\ref{Fig2}c. These emissions are not simply the bulk $sp$ bands scattered by the overlayer since mere scattering would replicate the Cu $sp$ band over the entire energy range. Instead one observes that they do not extend up to $E_F$ but stop at 1.2 eV binding energy, just below the onset of the LUMO orbital. We therefore suggest they are due to interfacial Cu $sp$ bands hybridizing with the 5A LUMO such that states appearing in these bands in the 0--1 eV binding energy range have changed their character from Cu $sp$ to that of the molecular LUMO.

\section{Conclusion}

In summary, a very strong substrate-induced dispersion of the LUMO orbital of 5A/Cu(110) along the Cu-row direction is traced back to a significant hybridization between the organic and the metallic states,  while the LUMO of 5A/Ag(110) exhibits only a minor intermolecular dispersion. 
By making use of the reciprocal relation between the structure of real space orbitals and features in the momentum maps, we are able to deduce a significant geometrical modification of the LUMO orbital upon adsorption on Cu(110),  while it remains essentially free-molecule like on Ag(110).  We believe that the main difference between the adsorption behavior of pentacene on these two surfaces arises from the distinct molecular orientation with respect to the close-packed metal rows. While on Ag(110) pentacene orients perpendicular to the rows, it aligns parallel to the metal rows on Cu(110). Because the electronic structure, e.g. the band dispersion of the metallic $sp$-bands or the location of the surface state in the surface Brillouin zone, is distinctly different along the rows $[1\overline{1}0]$ and perpendicular to them $[001]$, also the hybridization with adsorbed elongated species can be expected to be different.

In conclusion, for the case of pentacene monolayers on Ag(110) and Cu(110) surfaces, we have demonstrated the power of the orbital-tomography method using extensive angle-resolved photoemission data for revealing the electronic structure of such two-dimensional organic-metal interfacial layers. 
By generalizing the theoretical description of the photoemission process from isolated molecule systems to extended two-dimensional systems, we are able to explain the ''fine-structure'' in the experimental momentum maps and obtain a comprehensive understanding of the band dispersion of organic-metal interfacial layers.

\acknowledgements
We acknowledge financial support from the Austrian Science Fund (FWF) P21330-N20 and P23190-N16. We further acknowledge the Helmholtz-Zentrum Berlin -  Electron storage ring BESSY II for provision of synchrotron radiation at beamline U125/2-SGM.


%

\end{document}